\documentclass[aps,prl,twocolumn,groupedaddress]{revtex4-1}
\usepackage{graphicx}
\usepackage{subfigure}

\begin{document}

\title{Spin-asymmetric graphene nanoribbons in graphane on silicon dioxide}
\author{M. Ij\"{a}s$^1$}
\email{mari.ijas@aalto.fi}
\author{P. Havu$^1$}
\author{A. Harju$^1$}
\author{P. Pasanen$^2$}
\affiliation{$^1$Department of Applied Physics and Helsinki Institute of Physics, Aalto University, FI-02150 Espoo, Finland\\$^2$Nokia Research Center, P.O. Box 407, Helsinki FI-00045 Nokia Group, Finland}

\date{\today}

\begin{abstract}

Hydrogenated graphene, graphane, is studied on oxygen-terminated
silicon dioxide substrate using \emph{ab initio} calculations. The two lowest-energy structures with quarter and half mono-layer hydrogen coverage are presented.  We form zigzag graphene
nanoribbons by selectively removing hydrogens from the
epitaxial graphane layer. {In these ribbons, the spin degeneracy of the freestanding antiferromagnetic zigzag ribbons is broken, and band gaps of different magnitude emerge for the opposite spin species.} This degeneracy breaking is due to a charge imbalance in the substrate
below the ribbon, introduced through the asymmetric alignment of the
substrate atoms with respect to the edges of the graphene ribbon. {As the edge geometry is restricted by the neighboring graphane, the zigzag edges are robust to reconstructions suggested to destroy edge magnetism in freestanding graphene ribbons.}

\end{abstract}

\maketitle

Ever since the first experimental discovery of graphene and its
extraordinary electronic properties~\cite{Novoselov-Geim}, the search
for practical applications for this new material has been intense.  As
the lack of a band gap is a hindrance for many applications, various
methods have been suggested to induce a gap, among them cutting the
material into thin stripes, graphene nanoribbons (GNRs), that have a
width- and edge-dependent gap~\cite{Han}. In addition, due to the
antiferromagnetic spin alignment between the ribbon edges, zigzag
graphene nanoribbons (ZGNR) show potential as spin filters~\cite{Hod}. In the
presence of an external electric field across the ribbon width, they
are predicted to turn into half-metals~\cite{Son, Kan}, one of the
spin channels being insulating and the other metallic. The
halfmetallicity might be facilitated by different
edge-terminating groups~\cite{Hod, Li-Huang-Duan} and chemical doping
by boron and nitrogen~\cite{Dutta,Wu, Pruneda}.

The existence of a graphene derivative, the fully hydrogenated
graphane, was first theoretically predicted as late as in
2007~\cite{Sofo} and also experimentally verified soon after
that~\cite{Elias, Ryu}. Freestanding graphane, with hydrogenation on
both sides of the graphene layer, has been theoretically predicted to
be energetically stable~\cite{Sofo}, unlike one-sided
hydrogenation~\cite{Zhou-Wang, Zhou-Wu}. In the experimental
fabrication procedures, like exposure of a graphene layer deposited
on a substrate to atomic hydrogen~\cite{Ryu} or hydrogen
plasma~\cite{Elias}, it is unlikely that the hydrogenation
occurs on both sides.

For both graphene and graphane, the effect of a substrate has widely
been neglected in theoretical works, where focus has been on modeling
of freestanding graphene. Silicon dioxide is a likely substrate
candidate as it is insulating, cheap, and widely-used in the current
technologies.  The previous studies of graphene on SiO$_2$~\cite{Kang,
  Ao, Shemella-Nayak} are contradictory {whether or not graphene-substrate bonds are formed.} To the best of our knowledge, graphane
on a substrate has not been studied.  For simplicity, the one-sidedly
hydrogenated graphene layer on a substrate is called graphane in this work (also
the name "graphone" has been used in the literature \cite{Zhou-Wang}).

An interesting suggestion is that graphene nanoribbons could be drawn
in graphane by selective
dehydrogenation~\cite{Singh-Yakobson}. Dehydrogenation using an STM
tip has been experimentally
demonstrated~\cite{Sessi}. Theoretical
calculations~\cite{Singh-Yakobson, Hernandez-Nieves} predict that the
properties of graphene nanoribbons embedded in freestanding graphane
are very close to those of freestanding nanoribbons. The effect of the
substrate on such nanoribbons is, however, not known. {The main topic of our letter is to study graphene nanoribbons formed in graphane on
SiO$_2$.} We show that the ribbons show interesting electronic structure with spin-asymmetric bands due to the interaction with the
substrate. {Moreover, the neighboring graphane restricts edge reconstructions predicted \cite{Wassmann, Koskinen} to destroy edge magnetism in freestanding GNRs. }

The \emph{ab initio} calculations were performed using
density-functional theory with a van der Waals (vdW)
correction~\cite{Tkatchenko}, implemented in the all-electron FHI-aims
code~\cite{AIMS}. For computational details,
see Ref.~\cite{compdet}.  Silicon dioxide bulk in the $\alpha$-phase was
modeled using slabs whose thickness was
three unit cells corresponding to a layer of 15.8~\AA{}. As periodic
boundaries were used {for all directions}, vacuum layer of approximately 20~\AA{} was
placed between adjacent slabs.  A graphene layer with an optimized
unit cell volume at the distance of 2~\AA{} from the O-terminated
SiO$_2$ (0001) surface was placed on both sides of the {SiO$_2$ slab} and the
carbon {atoms} and three uppermost atoms of the substrate were relaxed. {The two interfaces on opposite sides of the slab were equivalent. The mismatch between the lattice
vectors of graphene and the substrate was  small, only 1.3\%.} 

{To briefly compare our results for the graphene-silicon dioxide system}
to previous works~\cite{Kang, Ao, Shemella-Nayak}, we note that in our
calculations, graphene {does not form bonds with the substrate atoms.}  {This result is in agreement with calculations of Ao~\emph{et al.}~\cite{Ao} but} in contradiction to previous results by Kang
\emph{et al}~\cite{Kang} and those of Shemella and
Nayak~\cite{Shemella-Nayak}, who found {{carbon-oxygen bonds} in all cases studied except in the
case of a hydrogen-saturated surface with terminating OH groups}. The
neglection of vdW forces in their calculations does not explain this
discrepancy, as vdW provides additional attraction. {However, during the structural relaxation we have observed geometries with fairly small residual forces that show
carbon-oxygen bonding, possibly explaining the discrepancy.}

\begin{figure}
{\includegraphics[width=0.49\columnwidth]{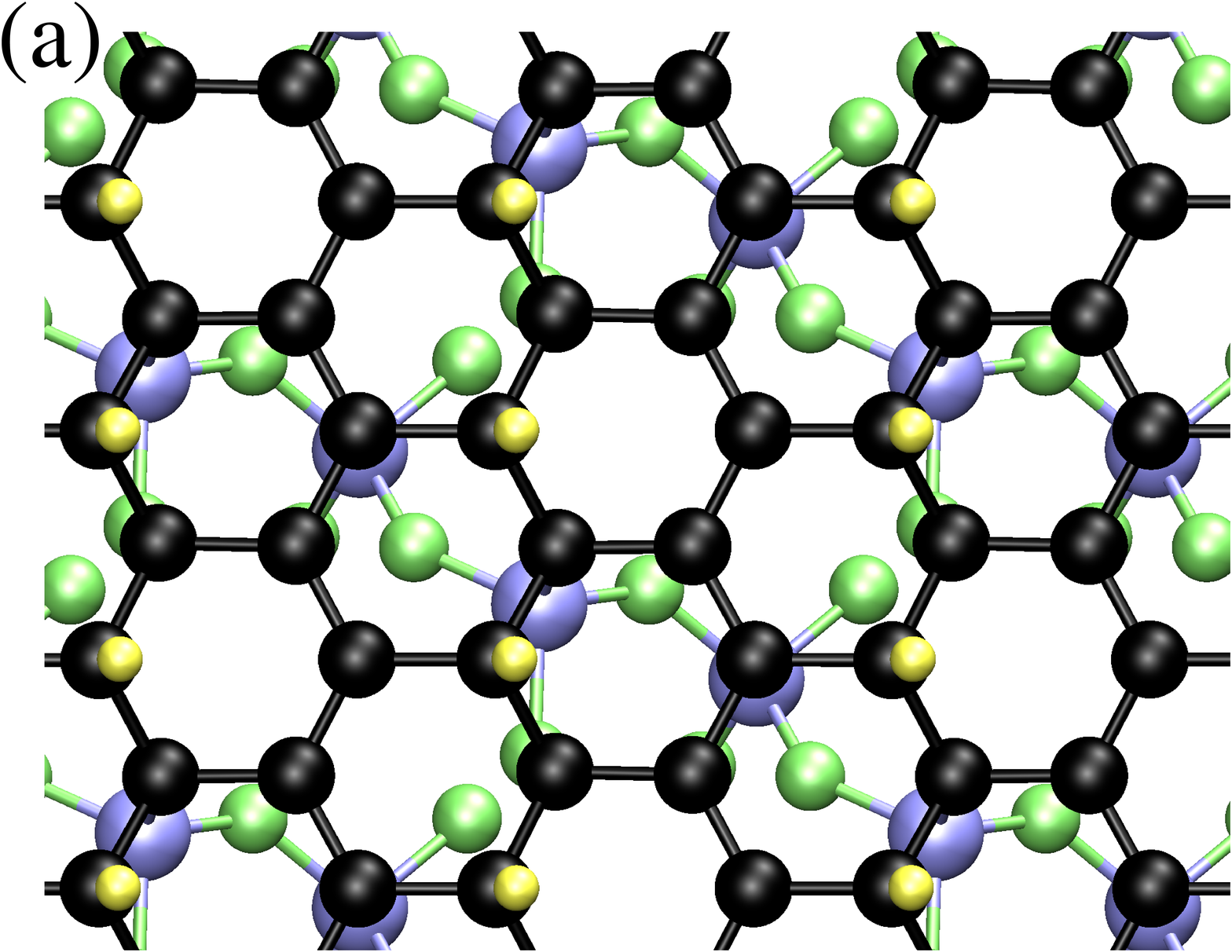}}
\hfill
{\includegraphics[width=0.49\columnwidth]{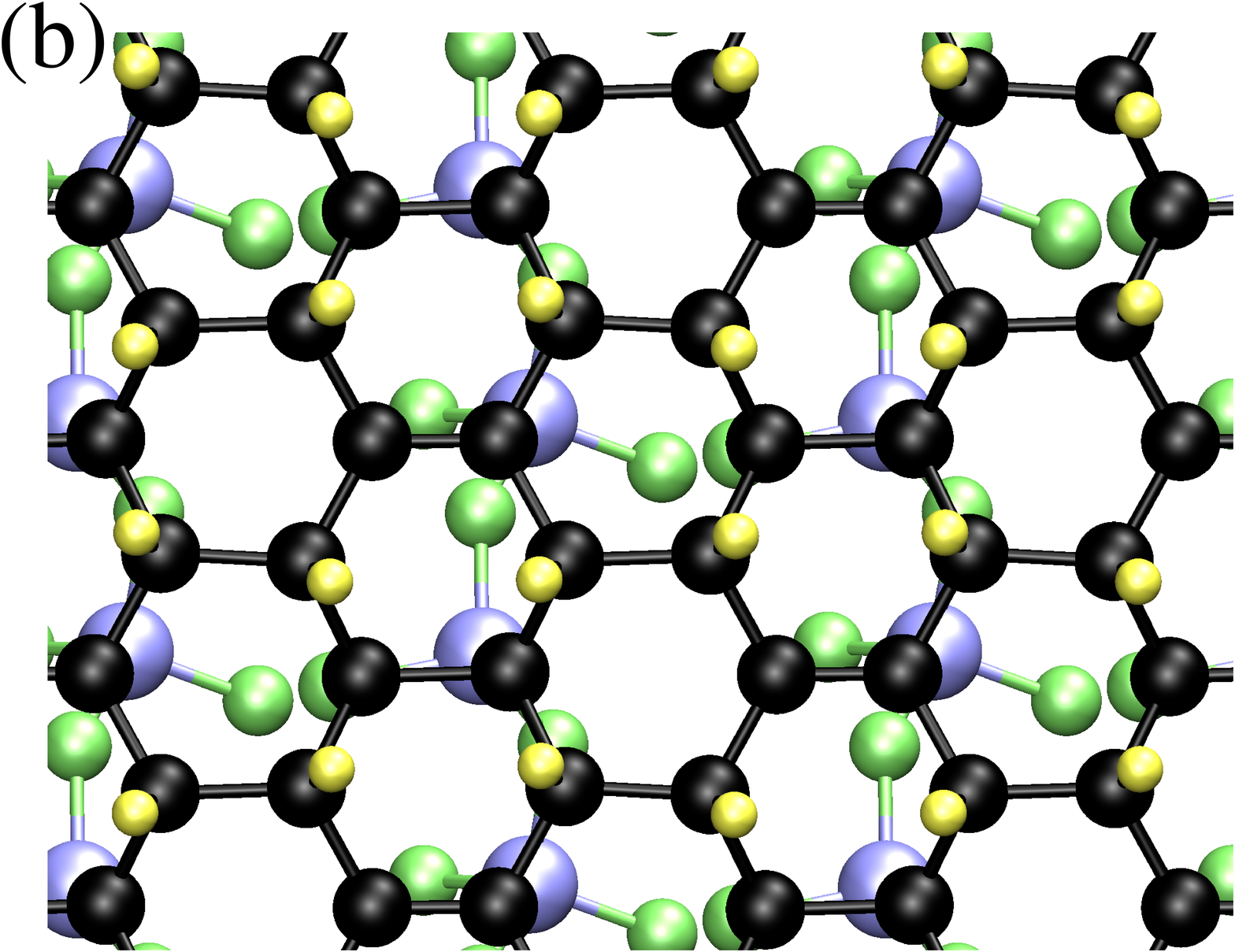}}\\
\caption{ \label{fig:graphane} 
(Color online) The two lowest-energy configurations for graphane on the
  O-terminated face of SiO$_2$. (a): The quarter-hydrogenated case
  (b): The half-hydrogenated case.
  Carbon atoms are black, other atoms from small to large: hydrogen, oxygen, and silicon }  
\end{figure}

For determining the lowest-energy hydrogen {atom} configurations on graphane,
one to six hydrogen atoms per unit cell containing eight carbon atoms
were placed initially either on the graphene carbon {atoms} or between
them{, corresponding to  1/6--3/4 hydrogen mono-layers (ML).}
{In scanning of a large number of structures, 
the SiO$_2$ layer was reduced to one unit cell}
corresponding to 5.5~\AA{}. The system was allowed relax the carbon
and hydrogen {atoms}. For the lowest-energy structures, calculations were
confirmed by using the thicker slab with three SiO$_2$ unit cells and
additionally relaxing three layers of the substrate atoms closest to
the surface. Altogether 492 initial hydrogen {atom} configurations were
considered. {A detailed analysis on the results is left for later work and we present here only results relevant for the nanoribbon calculations.}

The two most stable graphane configurations, with  {1/4~ML} and  {1/2~ML}
hydrogen coverage in the eight-carbon unit cell, are shown {in
  Fig.~\ref{fig:graphane}}. 
In the first configuration, the hydrogen {atoms} form lines along the zigzag chains
with a hydrogen {atom} on every second carbon atom and a free carbon chain
between two adjacent hydrogenated chains. On the hydrogenated rows,
the carbon {atoms} without hydrogen bind to the substrate.
In the second configuration, hydrogen {atoms} reside on all zigzag lines
but unlike in freestanding graphane hydrogenated on both sides, they
are on both sublattices, forming groups of four hydrogen {atoms} that are
slightly tilted towards each other. 
The two carbon {atoms} in the middle of each group
bind to a SiO-pair underneath.
The distance between the uppermost
substrate atoms and the closest atoms is almost the same for both
configurations, 1.41~\AA{} and 1.40~\AA{}, respectively. We could not
find a stable configuration for free-standing graphane that is
hydrogenated only on one side.

{Both configurations are {semiconductors} but their band gaps differ
considerably. The quarter-hydrogenated
configuration in {Fig.~\ref{fig:graphane}(a)} has only a minigap of 0.04~eV whereas the magnitude is 3.1~eV~for
half-hydrogenated graphane of {Fig.~\ref{fig:graphane}(b)}.} The
band gap of graphane has not been experimentally measured but for
graphane on iridium, a lower limit of 0.5~eV has been found \cite{Balog}. The gap calculated here is, of course, merely the
Kohn-Sham gap that is known to underestimate the real gap in LDA and
GGA calculations. The use of the GW self-energy correction could
improve the prediction and in the {double-sided} freestanding graphane the
band gaps increases by approximately
2~eV to 5.4-6.0~eV~\cite{Sahin, Lebegue}.

We choose the half-hydrogenated version as
a starting point for our ribbon simulations, {although the quarter-hydrogenated structure is lower in energy}. {Our calculations are performed at $T = 0$~K and in vacuum. Depending on the hydrogenation conditions, the more densely hydrogenated structure could be formed (see the thermodynamics-based analysis by Wassman \emph{et al.} \cite{Wassmann} on the stabilities of different GNR edge hydrogenations).} In the experiment, hydrogen {atoms} on graphene tend to cluster forming
denser structures~\cite{Balog} and we intend to model a ribbon drawn
through such an area.  Additionally, {the band gap in the half-hydrogenated
configuration is larger and its structure allows the formation of ribbons
with smoother edges}, in the fashion of earlier calculations on
freestanding graphene-graphane ribbons~\cite{Singh-Yakobson,
  Hernandez-Nieves}.

\begin{figure}
{  \includegraphics[width=.98\columnwidth]{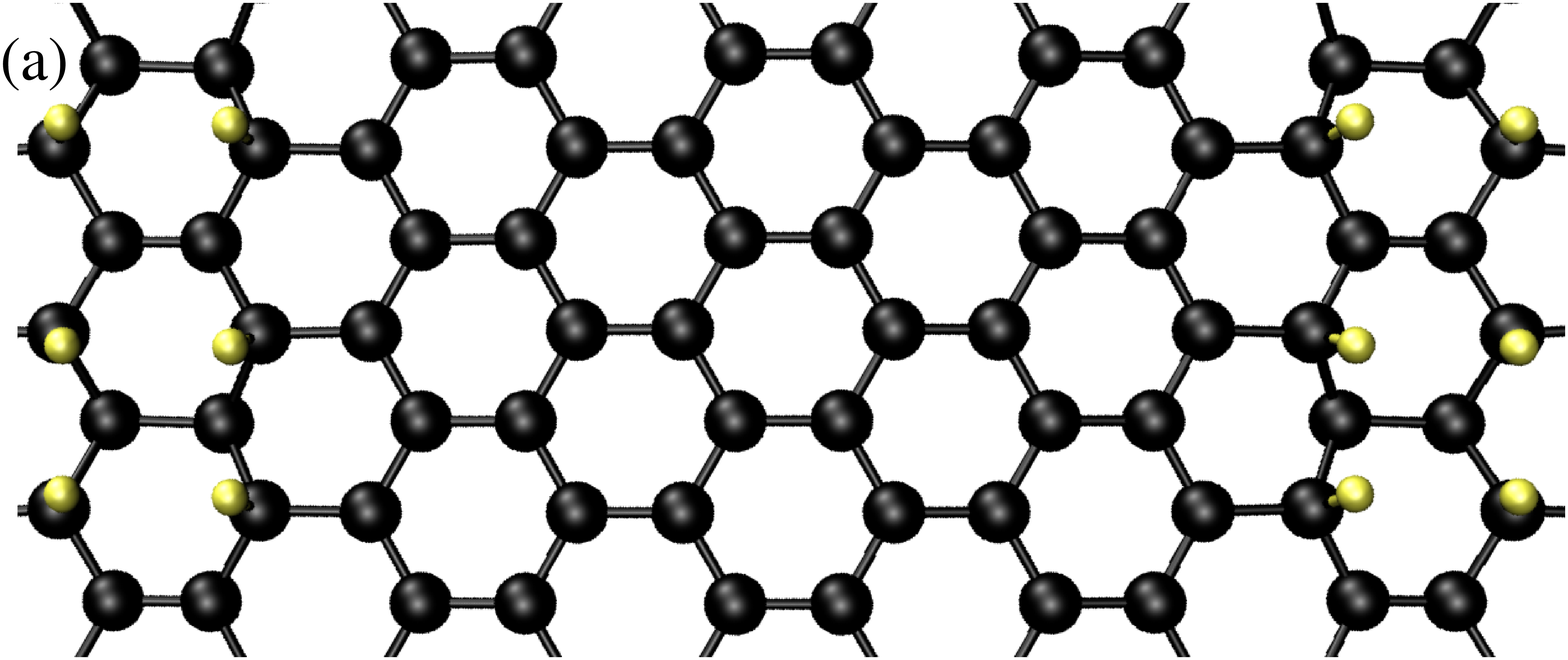}}
{  \includegraphics[width=.98\columnwidth]{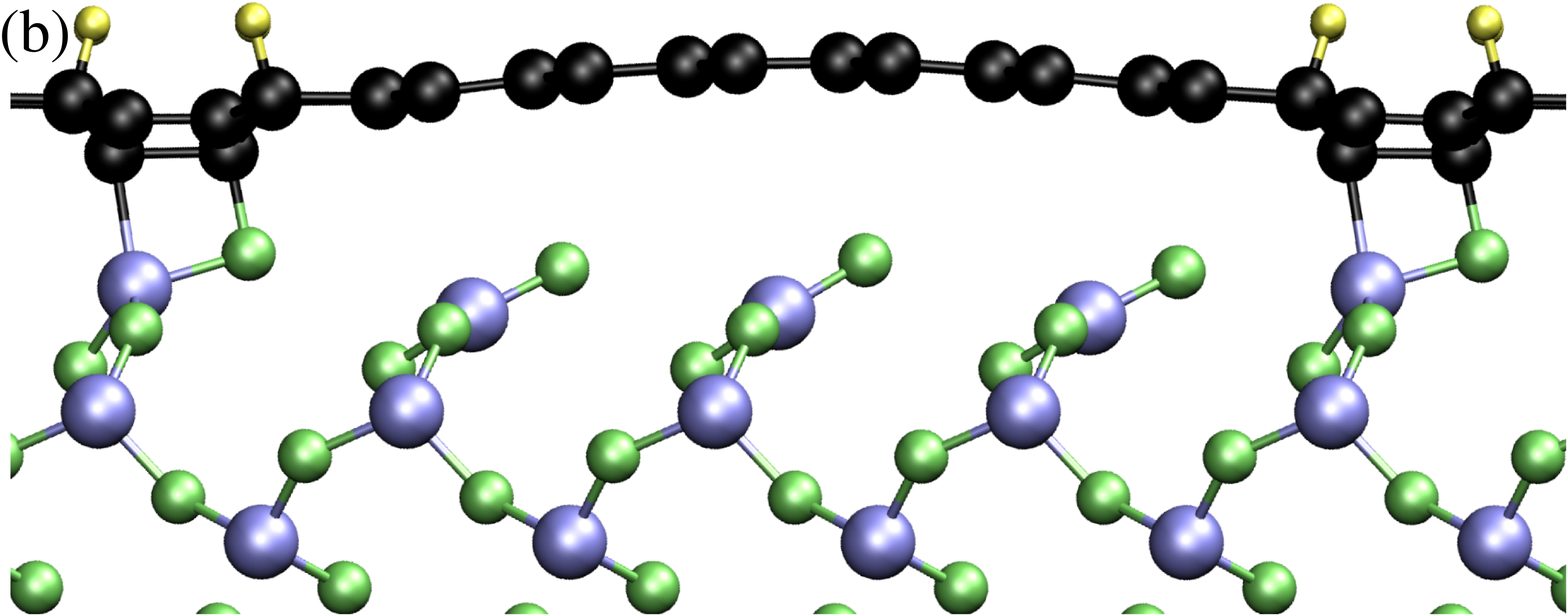}}
\caption{\label{fig:ribstruc} 
  (Color online) The six zigzag chains wide graphene ribbon in
  graphane on the SiO$_2$ substrate. {The visualization does not correspond to the calculational unit cell.} (a) Top view showing the hydrogen atom
  configuration with respect to the graphene layer, substrate atoms
  omitted for clarity. (b) Side view showing the bonding of graphane to substrate and the
  curvature of the graphene ribbon. The colors are as in Fig.~\ref{fig:graphane}.}
\end{figure}

The GNRs are formed by removing some of the hydrogen atoms from graphane, see Fig.~\ref{fig:ribstruc}(a). The ribbon is formed on one side of the
slab only, leaving the graphane layer on the other side intact. We have relaxed the
ribbon atoms, the CH rows on the edges of the ribbon, and the two
topmost {substrate atoms below it}. The
relaxation of further CH rows or substrate layers does not affect the
results. 
{The supercell used in the calculations was from four to nine graphane unit cells in the ribbon transverse direction in order to avoid interaction of the periodic images, the pristine graphane region between adjacent graphene strips
always being wider than the ribbon width.}
 Both antiferromagnetic (AFM) and ferromagnetic (FM) initial spin
 configurations were considered, and compared to the nonmagnetic
 case. In order to facilitate the convergence of the spin-polarized
 calculation, an AFM initial guess was obtained by using the spin
 configuration of a freestanding graphene ribbon in graphane. {As freestanding one-sided graphane was not found stable, double-sided graphane in the chair configuration was used.} {Ribbons
   with an odd number of zigzag lines were found to bind to
   the substrate on one of the edges, leading to localized states near
   the Fermi energy and absence of a ZGNR-like band structure.}

\begin{figure}
{\includegraphics[scale=0.36]{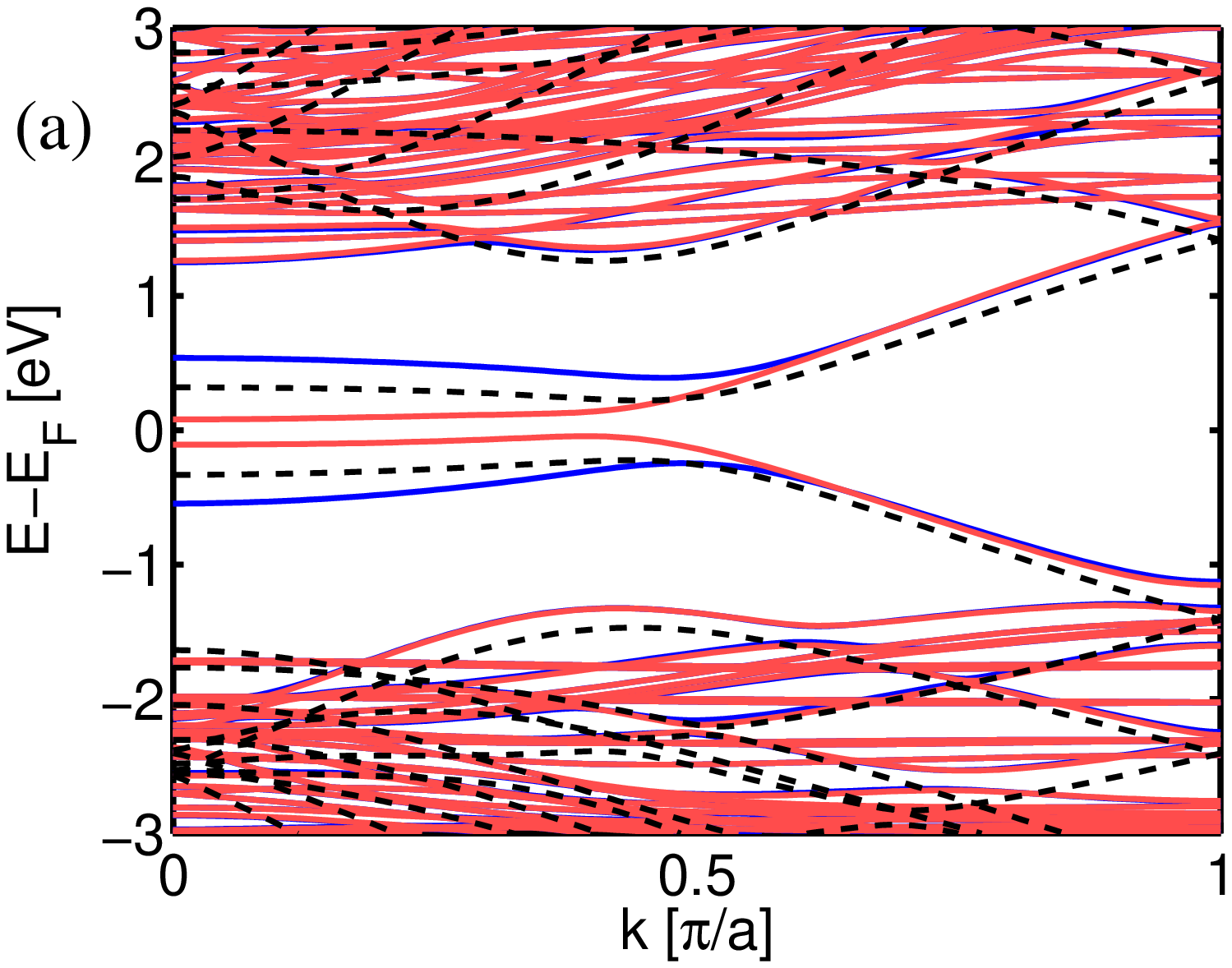}}
{\includegraphics[scale=0.36]{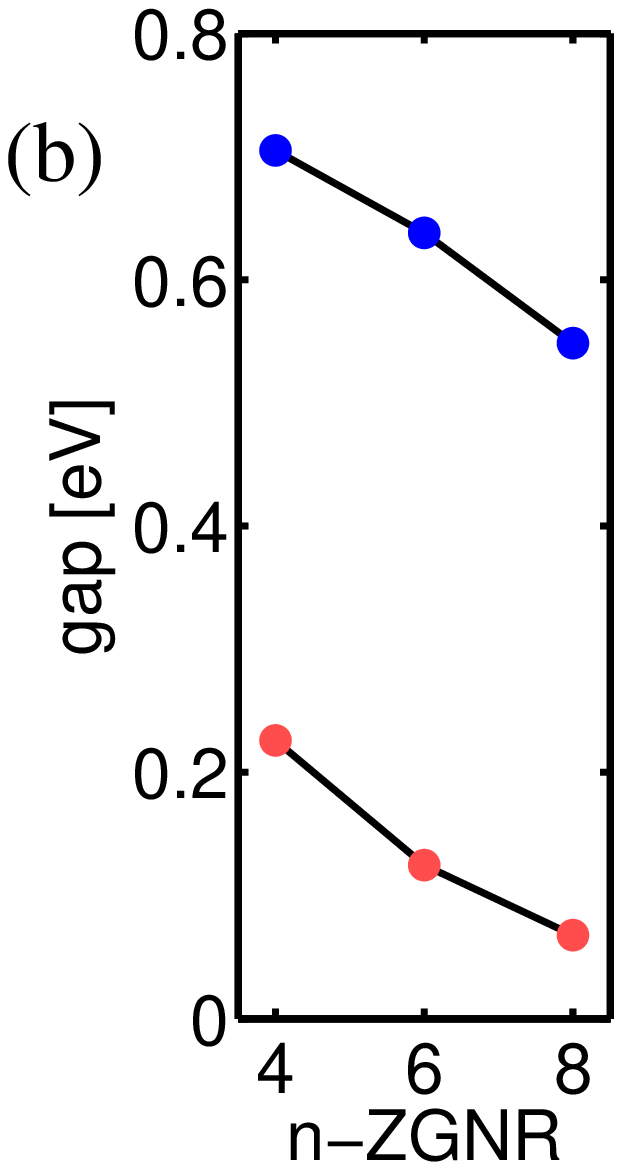}}
	
	\caption{\label{fig:bands} 
          (Color online) (a) The band structure of the antiferromagnetic
          6-ZGNR. The dashed lines show the spin-degenerate bands for
          the same ribbon in {double-sided} graphane without
          substrate. Blue (dark) and red (gray) correspond to the two
          spin orientations {that are non-degenerate only near $E_F$.} (b) The band gaps as a function of
          the ribbon width for both spin orientations.} 
\end{figure}

{The 4, 6, and 8 zigzag chains wide ribbons were studied and as a representative example, the relaxed structure of a six-chain
zigzag ribbon  is shown in Fig.~\ref{fig:ribstruc}.} For clarity, the substrate is removed
from the top view in Fig.~\ref{fig:ribstruc}(a).  Although the
carbon-carbon bond length in the ribbon is shorter than that of
graphane (1.441~\AA{} compared to 1.487~\AA{}, respectively) the
planarity of graphene together with the longer equilibrium distance
from the substrate causes the graphene ribbon to bend slightly, see
Fig.~\ref{fig:ribstruc}(b). {The narrow ribbons have more strain.} The
Si and O atoms below the ribbon are displaced by approximately
0.3-0.6~\AA{} from the positions corresponding to the optimized
SiO$_2$-graphane interface.

States with both AFM and FM spin order were found and like in the
absence of the substrate~\cite{Son}, the AFM was lower in energy. For
example, the energy difference between these two states was 25 meV per
edge carbon {atom} for 4-ZGNR {corresponding to a temperature of 290~K.} The nonmagnetic state was found to lie yet
higher in energy, the difference to the AFM state being 41 meV per
edge atom.

Fig.~\ref{fig:bands}(a) shows the band structure for the AFM 6-ZGNR. The
spin density within the ribbon area is analogous to that of
freestanding nanoribbons~\cite{Son}, {the two edges having opposite spin orientations and
the two sublattices being antiferromagnetically coupled}.  The hydrogen atoms at the interface are
spin-polarized to the same spin direction as the outermost carbon atoms of
the ribbon.  In Fig.~\ref{fig:bands}(a), also {the band structure for a GNR in free-standing double-sided graphane} is shown {in dashed lines}.  The substrate bands slightly
squeeze the ribbon bands together but, more importantly, in the
presence of the substrate, the spin degeneracy is broken and the spin
species have gaps of unequal magnitude. The change of the band
structure of a ZGNR towards half-metallicity has previously been
predicted to be induced by an external electric field~\cite{Son, Kan,
  Hod}, boron and nitrogen doping~\cite{Wu, Dutta} and asymmetric
terminating groups on the two edges of the ribbon~\cite{Hod,
  Li-Huang-Duan}. In this case, however, none of these factors is
present and the effect is given by the asymmetric coupling of the
ribbon edges to the substrate{, the inversion symmetry between the edges of the ribbon being broken.}  Although the highest occupied
Kohn-Sham orbitals are mostly localized at the ribbon, some density is
found around the uppermost oxygen atoms near the ribbon edge.  The
asymmetry of these oxygens leads then to the broken spin and charge
symmetries seen in the band structures. As some of the
charge density in the spin-polarized states has moved onto the
substrate, the ribbon {is} slightly doped. In
calculations using the Hubbard model, doping has been predicted to
lead to analogous band structures~\cite{Jung-MacDonald}.  The indirect
spin-polarized band gaps of the AFM ribbons decrease monotonically as
a function of the ribbon width, see Fig.~\ref{fig:bands}(b).

{To conclude, we have presented the two lowest-energy graphane configurations on oxygen-terminated silicon dioxide substrate with quarter and half hydrogen coverage of the graphene carbons and subsequently formed graphene nanoribbons in the half-hydrogenated structure.}
The presence
of the SiO$_2$ substrate has shown to substantially change properties
of the graphene nanoribbons embedded in graphane. More specifically, the asymmetry
introduced by the substrate structure breaks the spin degeneracy of
the antiferromagnetically ordered ribbons introducing spin-asymmetry in the gap.  Apart from that, these zigzag nanoribbons
resemble the freestanding ribbons showing antiferromagnetic spin order
between the ribbon edges and relatively flat bands close to the edge
of the Brillouin zone.   In
technological applications, such systems that could be turned to half-metallic could provide a spin-polarized current and function as spin filters. {Recently, the stability of zigzag-edged GNRs has been questioned as in freestanding ribbons, the edge is susceptible to reconstructions and hydrogen adsorption \cite{Koskinen, Wassmann}. In our system, the ribbon edge is fully saturated by graphane, and such reconstructions are unlikely. Thus, dehydrogenation could serve as a route to magnetism in GNRs.}

Finally, even if hydrogenating graphene evenly and in a stable manner
is currently experimentally challenging, very recent experimental and
theoretical studies \cite{Robinson, Zboril, Nair, Leenaerts} have
shown that stochiometrically fluorinated graphene is stable,
insulating, with properties similar to those of graphane. To our
knowledge, selective defluorination of graphene fluoride has not yet
been experimentally demonstrated. Theoretical calculations
\cite{Ribas} predict that the properties of ribbons obtained by
defluorination should be very similar to those obtained by
dehydrogenation. We thus expect the surface-induced
spin-asymmetry to be observable also in graphene
fluoride-based ribbons.

\emph{Acknowledgments---} 
We thank Ville Havu, Risto Nieminen, Andreas Uppstu, {Ilja Makkonen, and Martti Puska} for useful discussions. We acknowledge the support from 
Aalto-Nokia 
collaboration and Academy
of Finland through its Centers of Excellence Program (2006-2011). M. I. acknowledges the financial support from the Finnish Doctoral Programme in Computational Sciences FICS.

\bibliography{ribbons_prb}

\end{document}